\pdfoutput=1
\documentclass[12pt,a4paper]{article}

\usepackage{lineno}  % for line numbering during review
\usepackage{ifthen} % for conditional statements
\newboolean{pdflatex}
\setboolean{pdflatex}{true} % False for eps figures 
\newboolean{articletitles}
\setboolean{articletitles}{true} % False removes titles in references

\newboolean{uprightparticles}
\setboolean{uprightparticles}{false} %True for upright particle symbols
\usepackage{epstopdf}

\usepackage{xspace} % To avoid problems with missing or double spaces after
\usepackage{color}
\usepackage{colortbl}
\usepackage{amsmath} % Adds a large collection of math symbols
\usepackage{ifthen} % for conditional statements

\newboolean{inbibliography}
\setboolean{inbibliography}{true} %True once you enter the bibliography

\usepackage{longtable} % only for template; not usually to be used in PAPERs
\usepackage{graphicx}  % to include figures (can also use other packages)
\usepackage{epstopdf}

\usepackage{xspace} % To avoid problems with missing or double spaces after
\usepackage{color}
\usepackage{colortbl}
\usepackage{amsmath} % Adds a large collection of math symbols
\usepackage{ifthen} % for conditional statements

\usepackage{hyperref}    % Hyperlinks in references
\usepackage[all]{hypcap} % Internal hyperlinks to floats.
\usepackage{accents}

\def\lhcb {LHCb\xspace}
\def\ux85 {UX85\xspace}

\def\dzero  {D0\xspace}

\ifthenelse{\boolean{uprightparticles}}%
{

 \def\Pmu         {\ensuremath{\upmu}\xspace}

 \def\Ppi         {\ensuremath{\uppi}\xspace}

 \def\Ppsi        {\ensuremath{\uppsi}\xspace}

 \def\PDelta      {\ensuremath{\Delta}\xspace}                 
 \def\PXi      {\ensuremath{\Xi}\xspace}                 
 \def\PLambda      {\ensuremath{\Lambda}\xspace}                 
 \def\PSigma      {\ensuremath{\Sigma}\xspace}                 
 \def\POmega      {\ensuremath{\Omega}\xspace}                 
 \def\PUpsilon      {\ensuremath{\Upsilon}\xspace}                 
 
 \def\PB      {\ensuremath{\mathrm{B}}\xspace}                 
                  
 \def\PD      {\ensuremath{\mathrm{D}}\xspace}

 \def\PJ      {\ensuremath{\mathrm{J}}\xspace}                 
 \def\PK      {\ensuremath{\mathrm{K}}\xspace}

 \def\Pb      {\ensuremath{\mathrm{b}}\xspace}                 
 \def\Pc      {\ensuremath{\mathrm{c}}\xspace}                 
                  
 \def\Pe      {\ensuremath{\mathrm{e}}\xspace}

 \def\Pi      {\ensuremath{\mathrm{i}}\xspace}

 \def\Ps      {\ensuremath{\mathrm{s}}\xspace}

}
{

 \def\Pmu         {\ensuremath{\mu}\xspace}

 \def\Ppi         {\ensuremath{\pi}\xspace}

 \def\Ppsi        {\ensuremath{\psi}\xspace}                 
                  
 \mathchardef\PDelta="7101
 \mathchardef\PXi="7104
 \mathchardef\PLambda="7103
 \mathchardef\PSigma="7106
 \mathchardef\POmega="710A
 \mathchardef\PUpsilon="7107
                  
 \def\PB      {\ensuremath{B}\xspace}                 
                  
 \def\PD      {\ensuremath{D}\xspace}

 \def\PJ      {\ensuremath{J}\xspace}                 
 \def\PK      {\ensuremath{K}\xspace}

 \def\Pb      {\ensuremath{b}\xspace}                 
 \def\Pc      {\ensuremath{c}\xspace}                 
                  
 \def\Pe      {\ensuremath{e}\xspace}

 \def\Pi      {\ensuremath{i}\xspace}

 \def\Ps      {\ensuremath{s}\xspace}

}

\def\en         {\ensuremath{\Pe^-}\xspace}   % electron negative (\em is taken)
\def\ep         {\ensuremath{\Pe^+}\xspace}

\def\mup        {\ensuremath{\Pmu^+}\xspace}
\def\mun        {\ensuremath{\Pmu^-}\xspace} % muon negative (\mum is taken)

\def\squark    {\ensuremath{\Ps}\xspace}

\def\cquark    {\ensuremath{\Pc}\xspace}

\def\bquark    {\ensuremath{\Pb}\xspace}

\def\pion  {\ensuremath{\Ppi}\xspace}

\def\pip   {\ensuremath{\pion^+}\xspace}
\def\pim   {\ensuremath{\pion^-}\xspace}

\def\pipm  {\ensuremath{\pion^\pm}\xspace}

\def\kaon  {\ensuremath{\PK}\xspace}
  \def\Kbar  {\kern 0.2em\overline{\kern -0.2em \PK}{}\xspace}

\def\Kz    {\ensuremath{\kaon^0}\xspace}
\def\Kzb   {\ensuremath{\Kbar^0}\xspace}
\def\KzKzb {\ensuremath{\Kz \kern -0.16em \Kzb}\xspace}
\def\Kp    {\ensuremath{\kaon^+}\xspace}
\def\Km    {\ensuremath{\kaon^-}\xspace}

\def\KpKm  {\ensuremath{\Kp \kern -0.16em \Km}\xspace}
\def\KS    {\ensuremath{\kaon^0_{\rm\scriptscriptstyle S}}\xspace}

  \def\Dbar    {\kern 0.2em\overline{\kern -0.2em \PD}{}\xspace}
\def\D       {\ensuremath{\PD}\xspace}

\def\Dz      {\ensuremath{\D^0}\xspace}
\def\Dzb     {\ensuremath{\Dbar^0}\xspace}
\def\DzDzb   {\ensuremath{\Dz {\kern -0.16em \Dzb}}\xspace}
\def\Dp      {\ensuremath{\D^+}\xspace}
\def\Dm      {\ensuremath{\D^-}\xspace}

\def\DpDm    {\ensuremath{\Dp {\kern -0.16em \Dm}}\xspace}

\def\Dstarp  {\ensuremath{\D^{*+}}\xspace}

\def\Ds      {\ensuremath{\D^+_\squark}\xspace}
\def\Dsp     {\ensuremath{\D^+_\squark}\xspace}
\def\Dsm     {\ensuremath{\D^-_\squark}\xspace}
\def\Dspm    {\ensuremath{\D^{\pm}_\squark}\xspace}

\def\B       {\ensuremath{\PB}\xspace}
  \def\Bbar    {\kern 0.18em\overline{\kern -0.18em \PB}{}\xspace}
\def\Bb      {\ensuremath{\Bbar}\xspace}
 
\def\Bz      {\ensuremath{\B^0}\xspace}

\def\Bu      {\ensuremath{\B^+}\xspace}
\def\Bub     {\ensuremath{\B^-}\xspace}
\def\Bp      {\ensuremath{\Bu}\xspace}
\def\Bm      {\ensuremath{\Bub}\xspace}

\def\Bs      {\ensuremath{\B^0_\squark}\xspace}
\def\Bsb     {\ensuremath{\Bbar^0_\squark}\xspace}

\def\jpsi     {\ensuremath{{\PJ\mskip -3mu/\mskip -2mu\Ppsi\mskip 2mu}}\xspace}

  \def\Y#1S{\ensuremath{\PUpsilon{(#1S)}}\xspace}% no space before {...}!

\def\L {\ensuremath{\PLambda}\xspace}
\def\Lbar{\ensuremath{\overline \L}\xspace}

\def\Lb      {\ensuremath{\L_\bquark}\xspace}
\def\Lbbar   {\ensuremath{\Lbar_\bquark}\xspace}

\def\to                 {\ensuremath{\rightarrow}\xspace}

\def\CP                {\ensuremath{C\!P}\xspace}

\def\AT#1     {\ensuremath{A_{\mathrm{T}}^{#1}}\xspace}           % 2

\def\C#1      {\ensuremath{\mathcal{C}_{#1}}\xspace}                       % 9
\def\Cp#1     {\ensuremath{\mathcal{C}_{#1}^{'}}\xspace}                    % 7
\def\Ceff#1   {\ensuremath{\mathcal{C}_{#1}^{\mathrm{(eff)}}}\xspace}        % 9  
\def\Cpeff#1  {\ensuremath{\mathcal{C}_{#1}^{'\mathrm{(eff)}}}\xspace}       % 7
\def\Ope#1    {\ensuremath{\mathcal{O}_{#1}}\xspace}                       % 2
\def\Opep#1   {\ensuremath{\mathcal{O}_{#1}^{'}}\xspace}                    % 7

\newcommand{\tev}{\ensuremath{\mathrm{\,Te\kern -0.1em V}}\xspace}
\newcommand{\gev}{\ensuremath{\mathrm{\,Ge\kern -0.1em V}}\xspace}
\newcommand{\mev}{\ensuremath{\mathrm{\,Me\kern -0.1em V}}\xspace}
\newcommand{\kev}{\ensuremath{\mathrm{\,ke\kern -0.1em V}}\xspace}
\newcommand{\ev}{\ensuremath{\mathrm{\,e\kern -0.1em V}}\xspace}
\newcommand{\gevc}{\ensuremath{{\mathrm{\,Ge\kern -0.1em V\!/}c}}\xspace}
\newcommand{\mevc}{\ensuremath{{\mathrm{\,Me\kern -0.1em V\!/}c}}\xspace}
\newcommand{\gevcc}{\ensuremath{{\mathrm{\,Ge\kern -0.1em V\!/}c^2}}\xspace}
\newcommand{\gevgevcccc}{\ensuremath{{\mathrm{\,Ge\kern -0.1em V^2\!/}c^4}}\xspace}
\newcommand{\mevcc}{\ensuremath{{\mathrm{\,Me\kern -0.1em V\!/}c^2}}\xspace}

\def\invfb   {\ensuremath{\mbox{\,fb}^{-1}}\xspace}

\def\gsim{{~\raise.15em\hbox{$>$}\kern-.85em
          \lower.35em\hbox{$\sim$}~}\xspace}
\def\lsim{{~\raise.15em\hbox{$<$}\kern-.85em
          \lower.35em\hbox{$\sim$}~}\xspace}

\def\pt         {\mbox{$p_{\rm T}$}\xspace}

\def\tell1  {TELL1\xspace}
\def\ukl1   {UKL1\xspace}

\newcommand{\ptphi}{$p_{\rm T}\phi$}

\newcommand{\pxpy}{$p_xp_y$}

\newcommand{\asl}{$a_{\rm sl}^{\rm s}$}

\newcommand{\Ameas}{A_{\rm meas}}

\usepackage{mciteplus}
\usepackage{subfigure}
\begin{document}
%%%%%%%%%%%%%%%%%%%%%%%%%
%%%%% Title     %%%%%%%%%
%%%%%%%%%%%%%%%%%%%%%%%%%
\renewcommand{\thefootnote}{\fnsymbol{footnote}}
\setcounter{footnote}{1}

\begin{titlepage}
\pagenumbering{roman}
% Header ---------------------------------------------------
\vspace*{-1.5cm}
\centerline{\large EUROPEAN ORGANIZATION FOR NUCLEAR RESEARCH (CERN)}
\vspace*{1.5cm}
\hspace*{-0.5cm}
\begin{tabular*}{\linewidth}{lc@{\extracolsep{\fill}}r}
\ifthenelse{\boolean{pdflatex}}% Logo format choice
{\vspace*{-2.7cm}\mbox{\!\!\!\includegraphics[width=.14\textwidth]{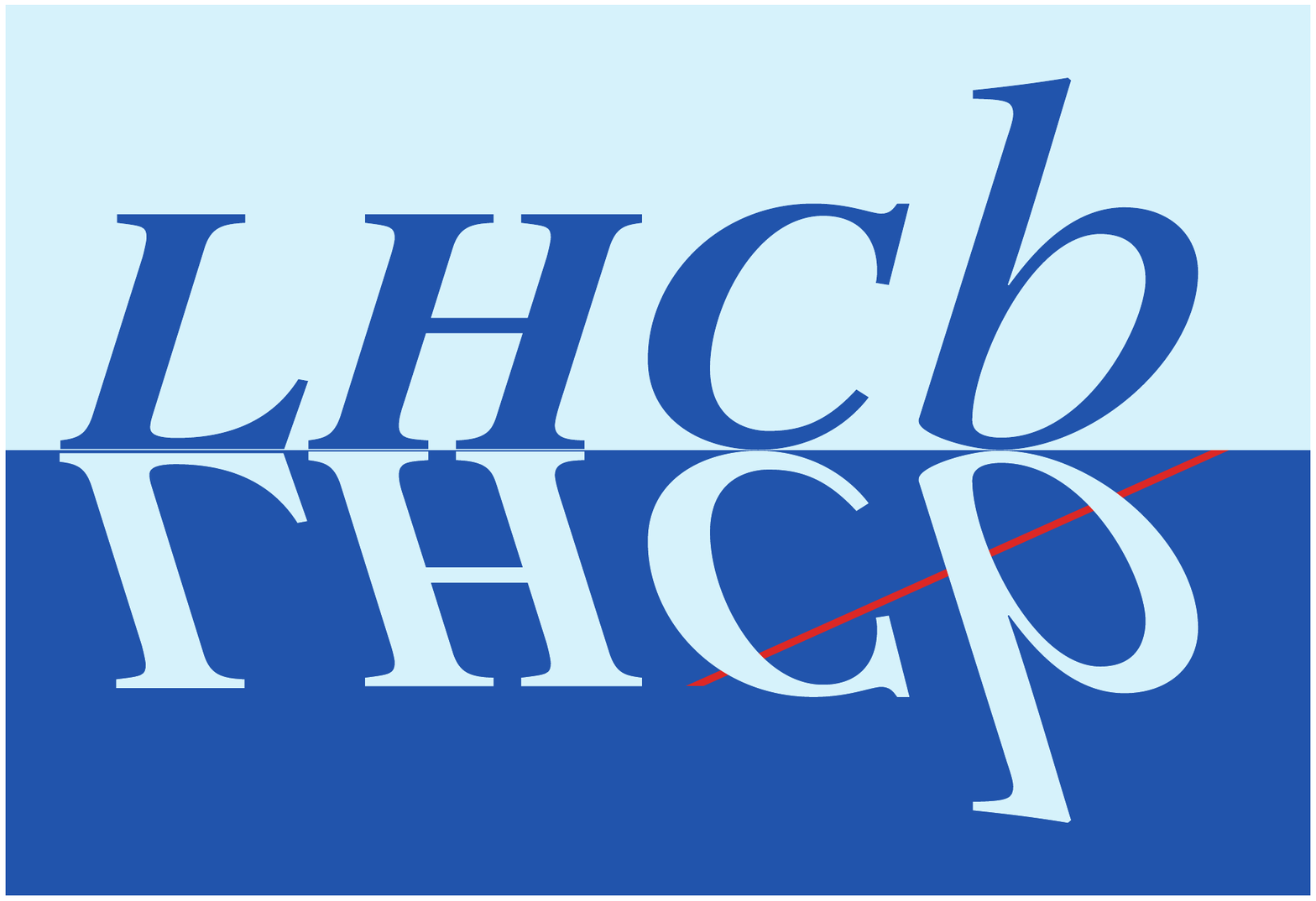}} & &}%
{\vspace*{-1.2cm}\mbox{\!\!\!\includegraphics[width=.12\textwidth]{newfigs/lhcb-logo.eps}} & &}%
\\
 & & CERN-PH-EP-2013-141 \\  % ID 
 & & LHCb-PAPER-2013-033 \\  % ID 
 & & October  25, 2013\\ % Date - Can also hardwire e.g.: 23 March 2010
 & & \\
% not in paper \hline
\end{tabular*}

%\vspace*{4.0cm}

\vspace*{4.0cm}

% Title --------------------------------------------------
\title{\bf\boldmath\huge
\begin{center}
Measurement of the flavour-specific  {\boldmath \CP }-violating asymmetry $a_{\rm sl}^s$ in \Bs decays 
\end{center}
}

\vspace*{2.0cm}

% Authors -------------------------------------------------
\begin{center}
The LHCb collaboration\footnote{Authors are listed on the following pages.}
\end{center}

%%%%%%%%%%%%%%%%%%%%%%%%%%%%%%%%%%%%%%%%%%

%\vspace{\fill}
%\clearpage

% Abstract -----------------------------------------------
\begin{abstract}
  \noindent
The \CP-violating asymmetry \asl\ is studied using  semileptonic decays of  $\Bs$ and $\Bsb$ mesons produced in $pp$ collisions at a centre-of-mass energy of 7~TeV at the LHC, exploiting a data sample corresponding to an integrated luminosity of 1.0\,fb$^{-1}$.  The reconstructed final states are  $\Dspm\mu ^{\mp}$, with the $\Dspm$ particle decaying in the $\phi\pipm$ mode. The  $\Dspm\mu ^{\mp}$ yields are summed over  $\Bsb$ and $\Bs$ initial states, and integrated with respect to decay time.
Data-driven methods are used to measure efficiency ratios. We obtain \asl\  = $(-0.06\pm0.50\pm0.36)$\%, where the first uncertainty is statistical and the second systematic.  
 \end{abstract}
\maketitle
%\vspace*{2.0cm}

\begin{center}
  Published in  Phys.~Lett.~B 
    \end{center}

\vspace{\fill}

{\footnotesize 
\centerline{\copyright~CERN on behalf of the \lhcb collaboration, license \href{http://creativecommons.org/licenses/by/3.0/}{CC-BY-3.0}.}}
\vspace*{2mm}

\end{titlepage}

%%%%%%%%%%%%%%%%%%%%%%%%%%%%%%%%
%%%%%  EOD OF TITLE PAGE  %%%%%%
%%%%%%%%%%%%%%%%%%%%%%%%%%%%%%%%

%  empty page follows the title page ----
\newpage
\setcounter{page}{1}
\mbox{~}
\newpage

% Author List ----------------------------
%  You need to get a new author list!
%\input{LHCb_authorlist.tex}
%%%%%%%%%%%%%%%%%%%%%%%%%%%%%%%%%%%%%%%%%%
\centerline{\large\bf LHCb collaboration}
\begin{flushleft}
\small
R.~Aaij$^{40}$, 
B.~Adeva$^{36}$, 
M.~Adinolfi$^{45}$, 
C.~Adrover$^{6}$, 
A.~Affolder$^{51}$, 
Z.~Ajaltouni$^{5}$, 
J.~Albrecht$^{9}$, 
F.~Alessio$^{37}$, 
M.~Alexander$^{50}$, 
S.~Ali$^{40}$
% ...

{\footnotesize \it
% ...
$ ^{5}$Clermont Universit\'{e}, Universit\'{e} Blaise Pascal, CNRS/IN2P3, LPC, Clermont-Ferrand, France\\
$ ^{6}$CPPM, Aix-Marseille Universit\'{e}, CNRS/IN2P3, Marseille, France\\
% ...
$ ^{9}$Fakult\"{a}t Physik, Technische Universit\"{a}t Dortmund, Dortmund, Germany\\
% ...
$ ^{36}$Universidad de Santiago de Compostela, Santiago de Compostela, Spain\\
$ ^{37}$European Organization for Nuclear Research (CERN), Geneva, Switzerland\\
% ...
$ ^{40}$Nikhef National Institute for Subatomic Physics, Amsterdam, The Netherlands\\
% ...
$ ^{45}$H.H. Wills Physics Laboratory, University of Bristol, Bristol, United Kingdom\\
% ...
$ ^{50}$School of Physics and Astronomy, University of Glasgow, Glasgow, United Kingdom\\
$ ^{51}$Oliver Lodge Laboratory, University of Liverpool, Liverpool, United Kingdom\\
% ...
}
\end{flushleft}
%%%%%%%%%%%%%%%%%%%%%%%%%%%%%%%%%%%%%%%%%%

%%%%%%%%%%%%%%%%%%%%%%%%%%%%%%%%%%%%%%%%%%
%%%%%%%%%%%%%%%%%%%%%%%%%%%%%%%%%%%%%%%%%%

\cleardoublepage

\renewcommand{\thefootnote}{\arabic{footnote}}
\setcounter{footnote}{0}

%%%%%%%%%%%%%%%%%%%%%%%%%%%%%%%%
%%%%%  Table of Content   %%%%%%
%%%%%%%%%%%%%%%%%%%%%%%%%%%%%%%%
%%%% Uncomment next 2 lines if desired
%\tableofcontents
\cleardoublepage

%%%%%%%%%%%%%%%%%%%%%%%%%
%%%%% Main text %%%%%%%%%
%%%%%%%%%%%%%%%%%%%%%%%%%

\pagestyle{plain} % restore page numbers for the main text
\setcounter{page}{1}
\pagenumbering{arabic}

%\linenumbers

\section{Introduction}

 The \CP asymmetry in $\Bs-\Bsb$ mixing  is a sensitive probe of new physics.
In the neutral $B$  system ($\Bz$ or $\Bs$), the mixing of the flavour eigenstates (the neutral $B$ and its antiparticle \Bb) is governed by a $2\times2$ complex effective Hamiltonian matrix \cite{Nierste:2009wg}
%\begin{widetext}
\begin{equation}
\begin{pmatrix}\noindent
M_{11}-\frac{i}{2}\Gamma_{11} & M_{12}-\frac{i}{2}\Gamma_{12} \\
M_{12}^*-\frac{i}{2}\Gamma_{12} ^* & M_{22}-\frac{i}{2}\Gamma_{22}
\end{pmatrix},
\end{equation}
%\end{widetext}
which operates on the neutral $B$ and $\Bb$  flavour eigenstates. The mass eigenstates have eigenvalues $M_{\rm H}$ and $M_{\rm L}$. Other measurable quantities are the mass difference $\Delta M$, the width difference 
$\Delta \Gamma$, and the semileptonic  (or flavour-specific) asymmetry $a_{\rm sl}$. These quantities are related to the off-diagonal matrix elements and the phase $\phi_{12}\equiv\arg\left(-M_{12}/\Gamma_{12}\right)$ by
\begin{eqnarray}
\Delta M&\equiv &M_{\rm H}-M_{\rm L} =2|M_{12}|\left(1-\frac{1}{8}\frac{|\Gamma_{12}|^2}{|M_{12}|^2}\sin^2\phi_{12}+....\right), \nonumber \\
\Delta \Gamma&\equiv&\Gamma_{\rm L}-\Gamma_{\rm H} =2|\Gamma_{12}|\cos\phi_{12}\left(1+\frac{1}{8}\frac{|\Gamma_{12}|^2}{|M_{12}|^2}\sin^2\phi_{12}+....\right), \nonumber \\
a_{\rm sl}&\equiv&\frac{\Gamma\left(\Bb(t)\to f\right)-\Gamma\left(B(t)\to \bar{f}\right)}{\Gamma\left(\Bb(t)\to f\right)+\Gamma\left(B(t)\to \bar{f}\right)}\simeq \frac{\Delta \Gamma}{\Delta M}\tan{\phi_{12}}\,,
\end{eqnarray}
where $B(t)$ is the state into which a produced $B$ meson has evolved after a proper time $t$ measured in the meson rest frame, and $f$ indicates a flavour-specific final state. The term flavour-specific means that the final state is  only reachable  by the  decay of the $B$ meson, and consequently reachable by a meson originally produced as a $\Bb$ only  through mixing.  We use the semileptonic  flavour specific final state and thus refer to this quantity as $a_{\rm sl}$. Note that $a_{\rm sl}$ is decay time independent.   Throughout the paper, mention of a specific channel implies the inclusion of the charge-conjugate mode, except in reference to asymmetries.

The phase $\phi_{12}$ is very small in the Standard Model (SM), in particular, for $\Bs$ mixing, $\phi_{12}^s$ is approximately $0.2^{\circ}$ \cite{Lenz:2006hd}.\footnote{This phase should not be confused with the \CP violation phase measured in $\Bs\to\jpsi\phi$ and $\Bs\to\jpsi\pi^+\pi^-$ decays, sometimes called $\phi_s$ \cite{Lenz-theory}.} New physics can affect this phase \cite{Bobeth:2011st,Lenz-theory} and therefore $a_{\rm sl}^s$.
The \dzero collaboration has reported evidence for a decay asymmetry  $A_{\rm sl}^b=(-0.787\pm0.172 \pm 0.093)\%$ in a mixture of \Bz and \Bs semileptonic decays, where the first uncertainty is statistical and the second systematic \cite{Abazov:2011yk,*Abazov:2010hv,*Abazov:2010hj}.  This asymmetry is much larger in magnitude than the SM predictions for semileptonic asymmetries in \Bs and \Bz decays, 
namely $a^s_{\rm sl}=(1.9\pm0.3)\times 10^{-5}$ and $a^d_{\rm sl}=(-4.1\pm 0.6)\times 10^{-4}$ \cite{Lenz-theory}.  More recently \dzero published measurements of $a^{d}_{\rm sl}= (0.68\pm0.45\pm0.14)\%$ \cite{Abazov:2012hha}, and $a^{s}_{\rm sl}=(-1.12\pm 0.74 \pm0.17)\%$ \cite{Abazov:2012zz}, consistent both with the anomalous asymmetry $A_{\rm sl}^b$ and the SM predictions for  $a^{s}_{\rm sl}$ and  $a^{d}_{\rm sl}$. If the measured value of $A_{\rm sl}^b$ is confirmed, this would demonstrate the presence of physics beyond the SM in the quark sector. The $\ep\en$ $B$-factory average asymmetry in \Bz decays is $a^d_{\rm sl}=(0.02\pm0.31)$\% \cite{Asner:2010qj}, in good agreement with the SM. A measurement of $a_{\rm sl}^s$ with comparable accuracy is important to establish whether physics beyond the SM influences flavour oscillations in the $\Bs$ system.

When measuring a semileptonic asymmetry at a $pp$ collider, such as the LHC, particle-antiparticle
production asymmetries, denoted as $a_{\rm P}$, as well as detector related asymmetries, may bias the measured value of $a_{\rm sl}^s$. We define $a_{\rm P}$ in terms of the numbers of produced $b$-hadrons, $N(B)$, and anti $b$-hadrons, $N(\Bb)$, as
\begin{equation}
a_{\rm P}\equiv \frac{N(B)-N(\overline{B})}{N(B)+N(\overline{B})}~,
\end{equation}
where $a_{\rm P}$ may in general be different for different species of $b$-hadron.

In this paper we report the measurement of  the asymmetry between $D_s^+X\mu^-\overline{\nu}$ and $D_s^-X\mu^+\nu$ decays, with $X$ representing  possible associated hadrons. We use the  $D_s^{\pm}\to \phi\pi^{\pm}$ decay. 
For a time-integrated measurement we have, to first order in \asl
\begin{equation}
\label{Eq:acceptrat}
A_{\rm meas}
\equiv \frac{\Gamma[{D_s^-\mu^+}]-\Gamma[{D_s^+\mu^-}]}{\Gamma[{D_s^-\mu^+}]+\Gamma[{D_s^+\mu^-}]}
=\frac{a_{\rm sl}^s}{2}+\left[a_{\rm P}-\frac{a_{\rm sl}^s}{2}\right]\frac{\int_{t=0}^{\infty} {e^{-\Gamma_s t}\cos ( \Delta M_s \, t )\epsilon(t)dt}}
{\int_{t=0}^{\infty} {e^{-\Gamma_s t}\cosh (\frac{\Delta\Gamma_s \, t}{2})\epsilon(t)dt}},
\end{equation}
where $\Delta M_s$ and $\Gamma _s$ are the mass difference and average decay width 
%the ``$s$" symbol specifies
of the $\Bs-\Bsb$ meson system, respectively, and $\epsilon(t)$ is the
decay time acceptance function for $\Bs$ mesons. Due to the large value of $\Delta M_s$, 17.768 $\pm $0.024 ps$^{-1}$ \cite{Aaij:2013mpa}, 
the oscillations are rapid and 
%It is given by 
%\begin{equation}
%\epsilon(t)=\frac{[1+\beta(t-t_0)][a(t-t_0)]^n}{1+[a(t-t_0)]^n},
%\end{equation}
%where $a=1.382$, $n=1.771$, $t_0=0.07742$ and $\beta=-0.0494$. 
the integral ratio in Eq.~(\ref{Eq:acceptrat})  is approximately 0.2\%.  Since the production asymmetry within the detector acceptance is expected to be at most a few percent \cite{Norrbin:2000jy,Aaij:2013iua,As:2012jw}, this reduces the effect of  $a_{\rm p}$ to the level of a few $10^{-4}$ for \Bs decays. This is well beneath our target uncertainty of the order of $10^{-3}$, and thus can be neglected, therefore yielding  $\Ameas$=0.5 \asl .

The measurement could be affected  by a
detection charge-asymmetry, which may be induced by  the event selection, tracking, and muon selection criteria.
The measured asymmetry can be written as
\begin{equation}
\Ameas =A_{\mu}^{\rm c}-A_{\rm track}-A_{\rm bkg},\label{eq:intasy}
\end{equation}
where $A_{\mu}^{\rm c}$  is given by
\begin{equation}
A_{\mu}^{\rm c}=\frac{N(\Dsm\mup)-N(\Dsp\mun)\times \frac{\epsilon(\mup)}{\epsilon(\mun)}}{N(\Dsm\mup)+N(\Dsp\mun)\times  \frac{\epsilon(\mup)}{\epsilon(\mun)}}.
\label{eq:amuc}
\end{equation}
$N(\Dsm\mup)$ and $N(\Dsp\mun)$ are the measured yields of $D_s \mu$ pairs, $\epsilon(\mup)$ and $\epsilon(\mun)$ are efficiency corrections  accounting for trigger  and muon identification effects, $A_{\rm track}$ is the  track-reconstruction asymmetry of charged particles, and $A_{\rm bkg}$ accounts for asymmetries induced by backgrounds. 
\section{The LHCb detector and trigger}
We use  a data sample corresponding to an integrated luminosity of  1.0\,\invfb  collected in 7 TeV $pp$ collisions with the LHCb detector \cite{Alves:2008zz}. This  detector is a single-arm forward
spectrometer covering the pseudorapidity range $2<\eta <5$, designed
for the study of particles containing \bquark or \cquark quarks. The
detector includes a high precision tracking system consisting of a
silicon-strip vertex detector surrounding the $pp$ interaction region,
a large-area silicon-strip detector located upstream of a dipole
magnet with a bending power of about $4{\rm\,Tm}$, and three stations
of silicon-strip detectors and straw drift-tubes placed
downstream. 
The combined tracking system has momentum resolution
$\Delta p/p$ that varies from 0.4\% at 5\gev to 0.6\% at 100\gev.\footnote{We work in units with $c$=1.} 
%and an impact parameter resolution of 20\mum for tracks with high
%transverse momentum, \pt. 
Charged hadrons are identified using two
ring-imaging Cherenkov (RICH) detectors~\cite{Adinolfi:2012an}. Photon, electron and hadron
candidates are identified by a calorimeter system consisting of
scintillating-pad and pre-shower detectors, an electromagnetic
calorimeter and a hadronic calorimeter. Muons are identified by a 
system composed of alternating layers of iron and multiwire
proportional chambers~\cite{Alves:2012ey}. The LHCb coordinate system is a right handed Cartesian system with the positive $z$-axis  aligned with the beam line and pointing away from the interaction point and the positive $x$-axis  following the ground of the experimental area, and pointing towards the outside of the LHC ring.

The trigger system~\cite{Aaij:2012me} consists of a hardware stage, based
on information from the calorimeter and muon systems, followed by a
software stage which applies a full event reconstruction. For the $D_s\mu$ signal samples, the hardware trigger (L0) requires the detection of a muon of either charge with  transverse momentum $\pt>1.64$\,GeV.  In
  the subsequent software trigger, a first selection algorithm  confirms the L0 candidate muon as a fully reconstructed track, while the second level algorithm  includes two possible selections. One is based on the topology of the candidate muon and one or two additional tracks,  requiring them to be detached from the primary interaction vertex. The second category is specifically designed to detect inclusive $\phi\to K^+K^-$ decays. We consider all  candidates that satisfy either  selection algorithm. We also study two mutually exclusive samples, one composed of candidates that satisfy  the second trigger category, and the other satisfying   the topological selection of  events including a muon, but not the inclusive $\phi$ algorithm.  Approximately 40\% of the data were taken with the magnetic field up, oriented along the positive $y$-axis in the LHCb coordinate system,  and the rest with the opposite down polarity. We exploit the fact that certain detection asymmetries cancel if data from different magnet polarities are combined.

\section{Selection requirements}
% Do not include this in analysis note and conference reports
%\input{acknowledgements}

%\input{appendix}

% This should be taken out in the final paper
%\input{supplementary-app}
Additional selection criteria exploiting the kinematic properties of semileptonic $b$-hadron decays \cite{Aaij:2010gn,Aaij:2011ju,Aaij:2011jp} are used.
In order to minimize backgrounds associated with misidentified muons, additional selection criteria on muons are  that the momentum, $p$, be between 6 and 100\,GeV,  that the pseudorapidity, $\eta$, be between 2 and 5, and that they are inconsistent with being produced at any primary vertex.  Tracks are considered as kaon candidates if they are identified by the RICH system,  have $\pt>0.3$\,GeV and  $p>2$\,GeV. The impact parameter (IP), defined as the minimum distance of approach of the track with respect to the primary vertex, is used to select tracks coming from charm decays. We require that the $\chi^2$, formed by using the hypothesis that each track's IP is equal to 0, which measures whether a track is consistent with coming from the PV, is greater than 9. 
  To be reconstructed as a $\phi$ meson candidate, a $K^+K^-$  pair must have  invariant mass within $\pm$20\,MeV of the $\phi$ meson mass. Candidate $\phi$ mesons are combined with charged pions to make $D_s$ meson candidates. The sum of  the \pt of $K^+$, $K^-$ and $\pi^{\pm}$ candidates must  be larger than 2.1\,GeV. The vertex fit $\chi^2$ divided by the number of degrees of freedom (ndf) must be less than 6, and the flight distance $\chi^2$, formed by using the hypothesis that the $\Ds$ flight distance is equal to 0, must be greater than 100. The \Bs candidate, formed from the $D_s$ and the muon, must have  vertex fit $\chi^2$/ndf $<6$, be downstream of the primary vertex, have $2< \eta< 5$ and have invariant mass between 3.1 and 5.1\,GeV. Finally, we include some angular selection criteria that require that the $B_s$ candidate have a momentum aligned with the measured fight direction. The cosine of the angle between the $D_s\mu$ momentum direction and the vector from the primary vertex to the $D_s\mu$ origin must be larger than 0.999.  The cosine of the angle between the $D_s$ momentum  and the vector from the primary vertex to the $D_s$ decay vertex must be larger than 0.99.

\section{Analysis method}
\begin{figure}[bt]
\begin{center}
\includegraphics[width=3 in]{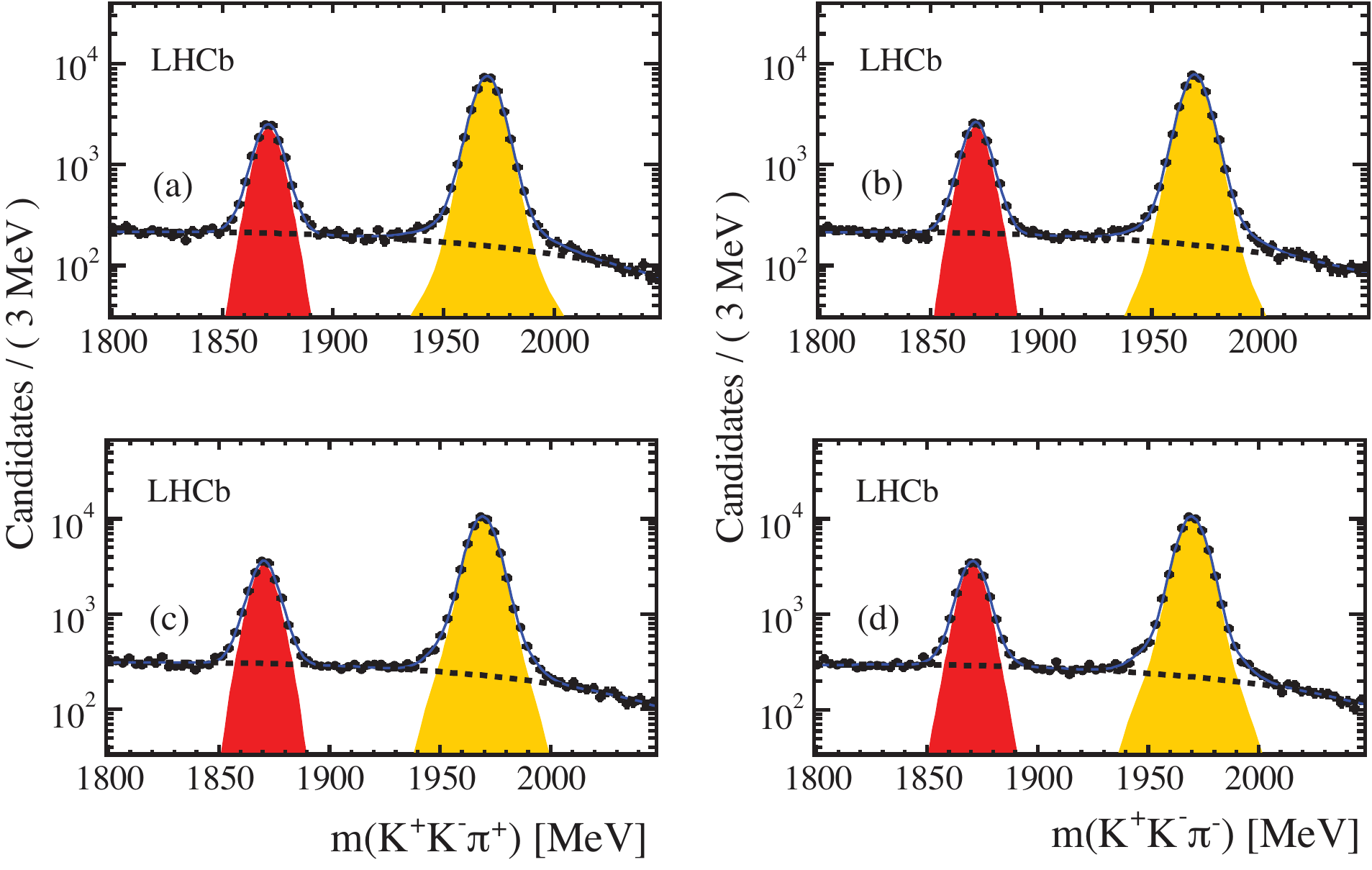}
\end{center}
\vspace{0pt}
\caption{\small Invariant mass distributions for: (a) $\Kp\Km\pip$  and (b) $\Kp\Km\pim$ 
candidates for magnet up, (c) $\Kp\Km\pip$ and (d) $\Kp\Km\pim$ candidates for magnet down with  $\Kp\Km$ invariant mass within $\pm$20 $\mev$ of the $\phi$ meson mass. The \Ds [yellow (grey) shaded area] and \Dp [red (dark) shaded area] signal shapes are described in the text. The $\chi ^2$/ndf   for these fits are 1.28, 1.25, 1.53, and 1.27 respectively, the corresponding p-values are
 7\%,  8\%, 4\%, 7\%.} 
\label{fig:Signal_mass_fit_up_after_fix}
\end{figure}

Signal yields are determined by fitting the $\Kp\Km\pi^+$ invariant mass distributions shown in 
Fig.~\ref{fig:Signal_mass_fit_up_after_fix}. 
We fit both the signal \Ds and \Dp peaks with double Gaussian  functions with  common means. The  \Dp\ channel  is used only as a component of the fit to the mass spectrum. The average mass resolution is about 7.1 \mev. The background is modelled with a second-order Chebychev polynomial.  The signal yields from the fits are listed in Table~\ref{tab:overall yields for Bs signal}. 
%As a check, the number of candidates are determined by counting the yields above background level in the range indicated by the vertical red-dashed lines above the background level,  determined in the fits discussed above. The signal yields are consistent using the two methods. 
%\clearpage

%The prompt background is of the order of 1\% of the total number of $\Ds$. Since the measured production asymmetry of $D_s^+ - D_s^-$ is $(-0.33\pm 0.22\pm 0.10)$\%  \cite{Aaij:2012cy}, the total effect on \asl\ is of the order of $3\times 10^{-5}$, and is negligible.  

%\clearpage

%The signal PDF of \Dspm is defined as:
%\begin{equation}
%f(m)=f_{1}Gauss(\mu_{1},\sigma_{1},m)+f_{2}Gauss(\mu_{2},\sigma_{2},m)+(1-f_{1}-f_{2})Gauss(\mu_{1},\sigma_{3},m)~,
%\end{equation} where $m$ indicates the observable, invariant mass $m(KK\pi)$, and $Gauss(\mu,\sigma,m)$ is the Gaussian function of $m$ with mean as $\mu$ and width as $\sigma$.
\begin{table}[t]
  \caption{\small  Yields for $\Dsp\mun$ and $\Dsm\mup$ events separately for magnet up and down data. These yields contain  very small contributions from  prompt $\D_s$ and b-hadron backgrounds. }
  \begin{center}\begin{tabular}{lcc}\hline
                     & magnet up        & magnet down       \\ 
      \hline
      $\Dsm\mup$         &  $38\,742\pm218$ & $53\,768\pm264$    \\
      $\Dsp\mun$         &  $38\,055\pm223$ & $54\,252\pm259$  \\
  \hline
  \end{tabular}\end{center}
  \label{tab:overall yields for Bs signal}
\end{table}

The detection asymmetry is largely induced by the dipole magnet, which bends particles of different charge in different detector halves.  The magnet polarity is reversed periodically, thus allowing the measurement and understanding of the size of this effect.  We analyze data taken with different magnet polarities separately, deriving charge asymmetry corrections for the two data sets independently. Finally, we average the two values in order to cancel charge any residual effects.
We  use two calibration samples containing muons to measure the relative trigger efficiencies of $D_s^+\mu^-/D_s^-\mu^+$ events, and the relative $\mu^-/\mu^+$ identification efficiencies. 
The first sample contains $b\to \jpsi(\to\mu^+\mu^-)X$ decays triggered independently of the \jpsi meson,  and where the $\jpsi$ is selected by requiring two particles  of opposite charge have an invariant mass consistent with the $\jpsi$ mass. This sample  is called the kinematically-selected (KS) sample. The second sample is collected by triggering on one muon from a \jpsi decay that is detached from the primary vertex. It is called muon selected (MS) as it relies on the presence of a well identified muon.

 In order to measure the relative $\pip$ and $\pim$ detection efficiencies,  we use the ratio of partially reconstructed and fully reconstructed $D^{*+}\to \pi^+D^0$, $D^0\to K^-\pi^+\pi^+(\pi^-)$ decays. The former sample is gathered without explicitly reconstructing the $\pi^-$ particle, and then the efficiency of finding this track in the event is measured. The same procedure is applied to the charge conjugate mode, so the relative $\pi^+$ to $\pi^-$ efficiency is measured.  A detailed description is given in Ref.~\cite{Aaij:2012cy}.

Finally, a  sample of $D^+(\to\Km\pip\pip)\mu^-$ candidates is obtained using similar triggers to the $D_s\mu$ sample.  This sample is used to assess charge asymmetries induced by the software trigger.
%\subsection{The measured semileptonic asymmetry}

%Detection asymmetries are largely induced by the dipole magnet, which bends particles of opposite charges
% into different detector halves.  The magnet polarity is reversed periodically, thus allowing the measurement and understanding of the size of this effect.  Data taken with different %magnet polarities are analyzed separately. Finally, we average the two results in order to cancel any residual effects.

\begin{figure}[t]
\begin{center}
\includegraphics[width=2.5 in]{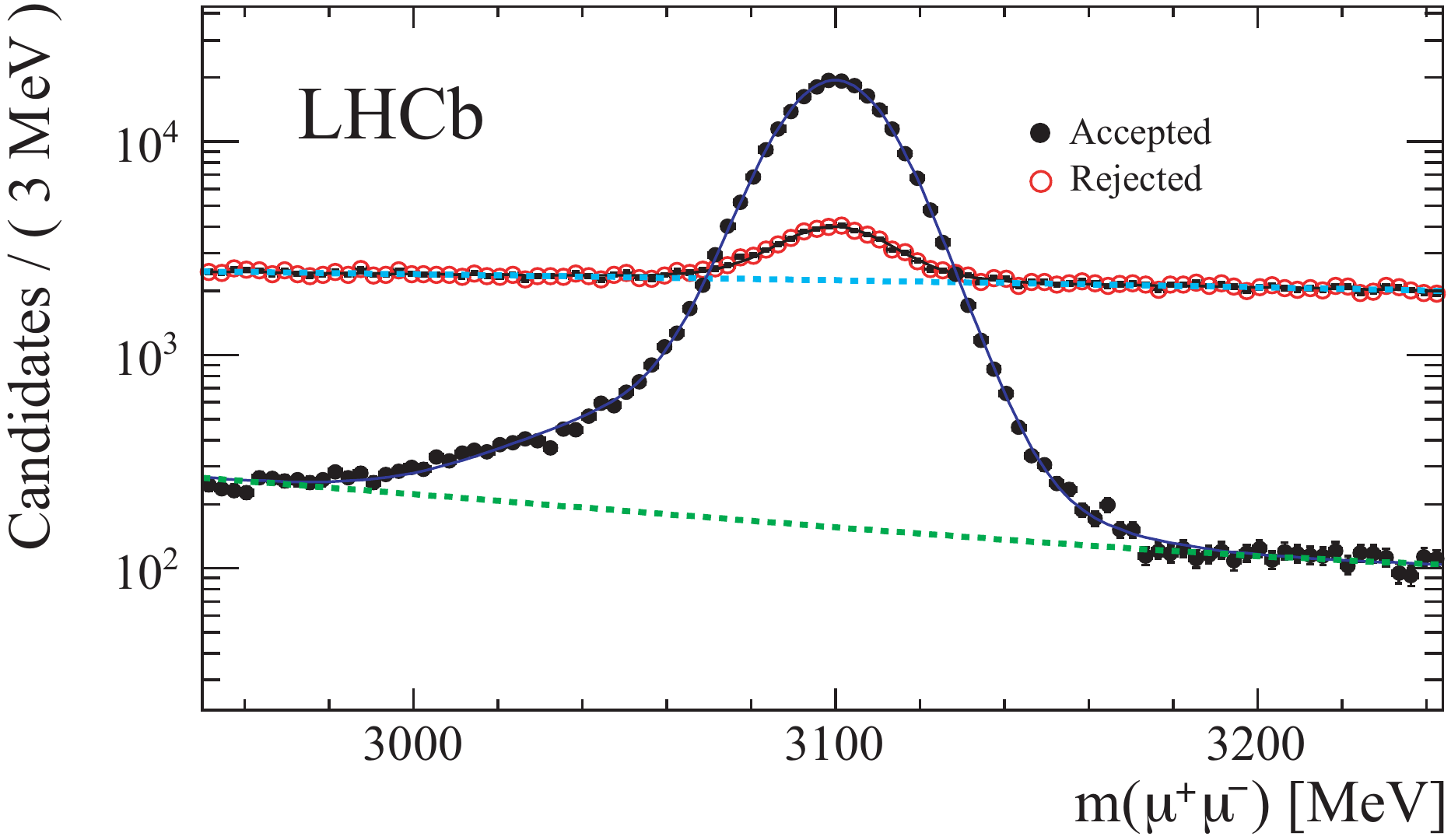}
\end{center}
\vspace{-6mm}
\caption{\small  Invariant $\mup\mun$ mass distributions of  the kinematically-selected \jpsi candidates in magnet up data, where the red (open) circles represent entries where the muon candidate, kinematically selected,  is rejected and the black (filled) circles those where it is accepted by the muon identification algorithm. The dashed lines represent the combinatorial background. } 
\label{fig:minbias_jpsi_mass_fit_mimic}
%The invariant mass distributions of $\mup\mun$ for the kinematically selected \jpsi events where we insist that one of the muons be completely independent of the trigger (called the %probe). In (a) the trigger efficiency is determined by comparing the number of well identified muon tracks accepted by the trigger, shown using the solid (black) circles, with the ones %rejected by the trigger shown in the (red) squares. In (b) the probe muon is used to determine the muon identification efficiency. This is magnet up data only, a similar sample is %used for magnet down data.} 
%\label{fig:minbias_jpsi_mass_fit}
\end{figure}

The efficiency ratio  $\epsilon_{\mup}/\epsilon_{\mun}$ in Eq.~(\ref{eq:amuc})  accounts for losses due to the muon identification efficiency algorithm and the
 trigger requirements. We measure $\epsilon_{\mup}/\epsilon_{\mun}$ using the KS and MS calibration samples. There are about 0.6 million  KS \jpsi candidates selected in total,  and about 1.2 million MS \jpsi candidates.
 As the calibration muon spectra are slightly softer than that of the signal,
 we subdivide the signal and calibration samples into subsamples defined by the kinematic properties of the candidate muon. We define five muon momentum bins: $6-20~\gev$, $20-30~\gev$, $30-40~\gev$, $40-50~\gev$, and  $50-100~\gev$. We further subdivide the signal and calibration samples with two binning schemes. In the first, each $\mu$ momentum bin is split into 10 rectangular regions in $qp_x$ and $p_y$, where $q$ represents the  muon charge and $p_x$ and $p_y$ are the Cartesian components of the muon momentum in the directions perpendicular to the beam axis. The second grid uses 8 regions of  muon  $p_{\rm T}$ and azimuthal angle $\phi$ to reduce the sensitivity to differences in $\phi$ acceptance between signal and calibration samples. In this case the first and third bins in $\phi$ are flipped for negative charges, to symmetrize the acceptance in a consistent manner with the $q p_x$ and $p_y$ binning. Signal  and calibration yields are determined separately in each of the intervals both for magnet up and down data. 
Figure~\ref{fig:minbias_jpsi_mass_fit_mimic} shows the $\mup\mun$ invariant mass distribution for the KS  \jpsi events in magnet up data.

%\subsection{Dipole magnet and acceptance effects}

\begin{figure}[t]
\begin{center}
\includegraphics[width=2.5 in]{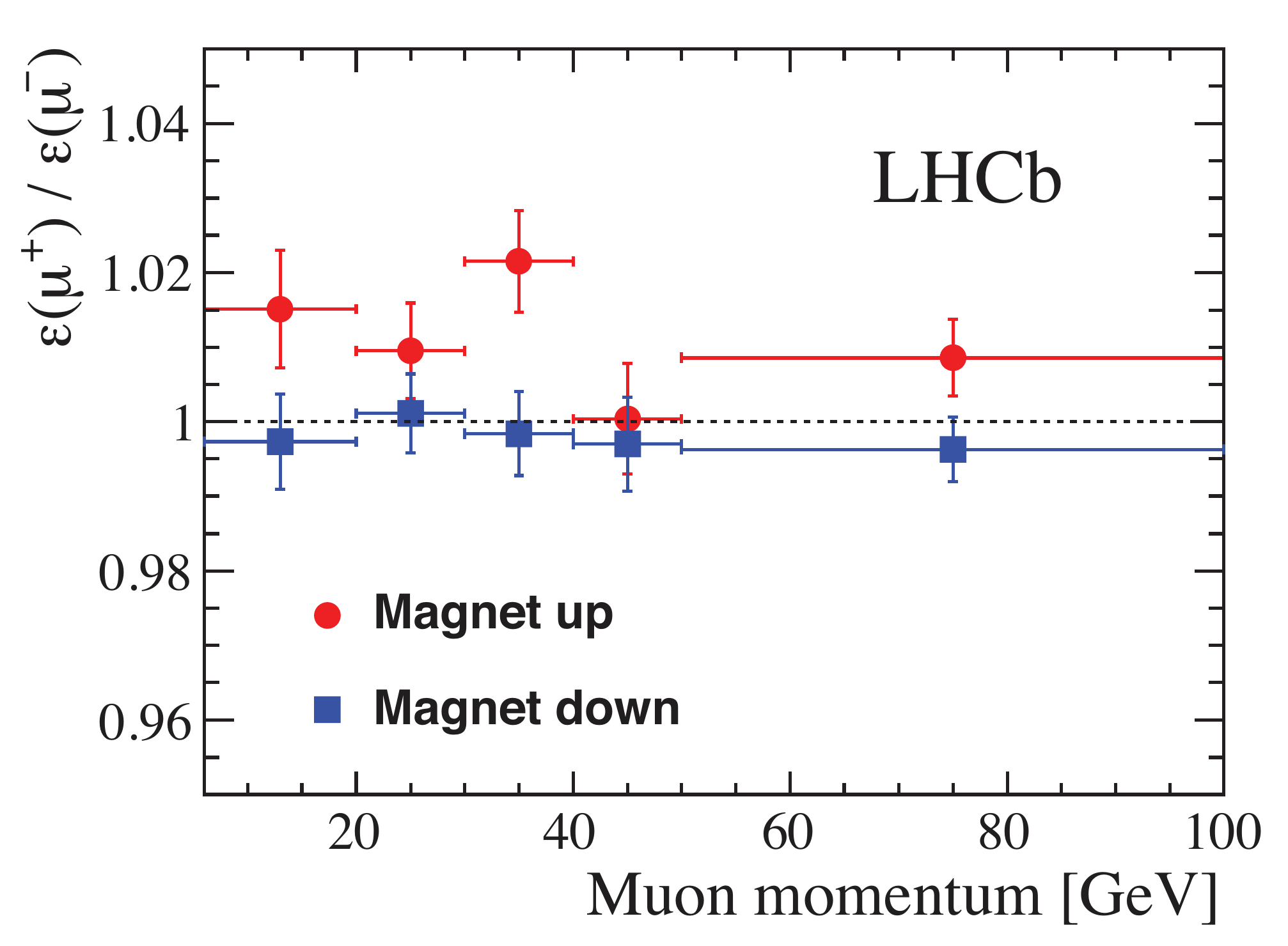}
\end{center}
\vspace{0pt}
\caption{\small  Relative muon efficiency as a function of muon  momentum determined using the kinematically-selected \jpsi\  sample.}
\label{fig: muon_relative_eff_p}
\end{figure}

The relative efficiencies for triggering and identifying muons in five different momentum bins are shown in Fig.~\ref{fig: muon_relative_eff_p} for magnet up and magnet down data using the KS calibration sample. They are consistent with being independent of momentum.  The small difference of approximately 1\% between the two samples can be attributed to the alignment of the muon stations, which affects predominantly the hardware muon trigger. 

%Figure~\ref{fig:Momentum_PID_MB_Sig} shows a comparison between the momentum spectra in the calibration samples and the data. The KS sample matches the signal better than the MS sample, however, it still differs from the signal sample.

The $D_s^+\mu^-$ final state benefits from several cancellations of potential instrumental asymmetries that can arise due to the different interaction cross-sections in the detector material or to differences between tracking reconstructions of negative and positive particles. The $\mu$ and $\pi$ charged tracks have very similar reconstruction efficiencies. 
Using the partially-reconstructed $D^{*+}$ calibration sample,
we found that the $\pi^+$ versus $\pi^-$ relative tracking efficiencies  are independent of momentum  and transverse momentum \cite{Aaij:2012cy}. This, along with the fact that $\pip$ and $\pim$ interaction cross-sections on isoscalar targets are equal, and that the detector is almost isoscalar,  implies that the difference between $\pip$ and $\pim$ tracking efficiencies depend only upon the magnetic field orientation and the detector acceptance. Thus the charge asymmetry ratios measured for  pions are applicable to  muons as well. In the $\phi\pi^+\mu^-$ final states, the pion and muon have opposite signs, and thus the charge asymmetry in the track reconstruction efficiency induced by imperfect $\pi\mu$ cancellation,~ $A_{\rm track}^{\pi\mu}$, is small. Using the efficiency ratios $\epsilon_{\pip}/\epsilon_{\pim}$ measured with the $D^{*+}$ calibration sample, we obtain $A_{\rm track}^{\pi\mu}=(+0.01\pm 0.13)$\%. A small residual sensitivity  to the charge asymmetry in $K$ track reconstruction   is present due to a slight momentum mismatch between the two kaons from $\phi$ decays arising from the interference with the S-wave component. It is determined to be $A_{\rm track}^{ KK}=(+0.012\pm0.004)$\%. The efficiency ratios used in determining $A_{\rm track}^{KK}$ are based on   $\epsilon_{\pip}/\epsilon_{\pim}$ with a correction derived from the comparison between the Cabibbo-favoured decays $\Dp\to \Km\pip\pim$ and $\Dsp \to \KS\pip$, accounting for additional charge asymmetry induced by $K$ interactions in the detector.
Therefore, the total tracking asymmetry is $A_{\rm track}=(+0.02\pm 0.13)$\%. %\clearpage

\section{Backgrounds}
\begin{figure}[t]
\begin{center}
\includegraphics[width=5in]{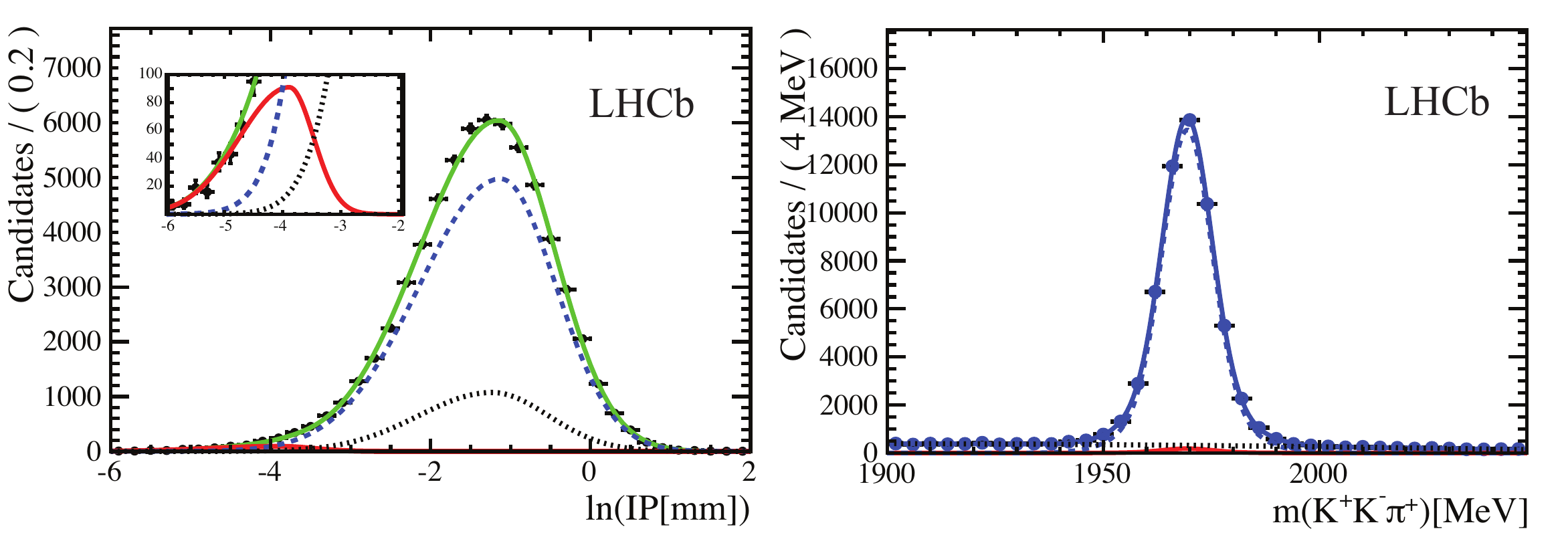}
\end{center}
\vspace{0pt}
\caption{\small (a) Spectrum of the logarithm of the IP calculated with respect to the primary vertex  for    \Dsp  candidates in combination with muons; the insert shows a magnified view of the region where the prompt $\Dsp$ contribution peaks. The blue dashed line is the component coming from $B$ hadron decays, the black dashed line the false \Ds background, the red line the prompt background, (b) the invariant mass distributions for $\Ds\to\phi\pi$ candidates. These distributions are for the magnet down sample. (For interpretation of the reference to colour in this figure legend, the reader is referred to the web version of this Letter.)}
\label{fig:Signal_2D_fit_down_DsplusMuminus_after_fix}
\end{figure}

Backgrounds include prompt charm production, fake muons associated with real $D_s^+$ particles produced in $b$-hadron decays, and  $B\to D D_s$ decays where the $D$ hadron decays semileptonically. Here $B$ denotes any meson or baryon containing a $b$ (or $\overline{b}$) quark, and similarly, $D$ denotes any hadron containing a $c$ (or $\overline{c}$) quark. 
The prompt background is highly suppressed by the requirement of a well identified muon forming a vertex with the  $\Ds$ candidate.  The prompt yield is separated from false $D_s$ backgrounds using a binned two-dimensional fit to the mass and ln(IP/mm) of the $\phi \pi^+$ candidates. The method is described in detail in Ref.~\cite{Aaij:2011jp}.
Figure~\ref{fig:Signal_2D_fit_down_DsplusMuminus_after_fix} shows the fit results for the magnet-down $D_s^+\mu^-$ candidate sample. From the asymmetry in the prompt yield normalized to the overall signal yield in the five momentum bins, we obtain an asymmetry due to prompt background equal to (+0.14$\pm$0.07)\% for magnet up data, ($-0.05\pm0.05)$\% for magnet down data, with an average value of (+0.04$\pm$0.04)\%.

%\subsection{\boldmath Fake muons in association with $\Bsb\to D_s^+ X$ decays}
 
Samples of $\Ds \pi^-X$ and $\Ds K^-X$ events, where $X$ represents undetected particles from the same decay, are used to infer the numbers of  \Ds-hadron combinations from $B$ decays that could be mistaken for  $\Ds \mu^-$ events if the hadron is misidentified as a muon. Kaons and pions are identified using the RICH.
These numbers, combined with knowledge of the probability that kaons or pions are mistaken for muons, provide a measurement of the fake hadron background.  These misidentification probabilities are also calculated in the five momentum bins using $\Dstarp\to\pip\Dz$ decays, with $\Dz$ decaying into the $\Km\pip$ final state. 
The net effect on the asymmetry is below $10^{-4}$ and thus the $\Ds$-hadron background can be ignored.

%\subsection{\boldmath Backgrounds from $B\to DD_s$ decays}
We also consider the background induced by  $D_s^+\mun$ events deriving from  $b\to c\bar{c}s$ decays where the $\Ds$ hadron originates from the virtual $W^+$ boson and the muon originates from the charmed-hadron semileptonic decay. These backgrounds are suppressed since the $D$ hadron travels away from the $B$ vertex prior to its semileptonic decay. As these decays are of opposite sign to the signal, they cause a background asymmetry that is proportional to the production asymmetry of the background sources. 
The $\Bz$ production asymmetry  has been measured in LHCb to be $(-0.1 \pm 1.0)$\%  \cite{Aaij:2013iua}, and the $\B^+$ production asymmetry to be  $(+0.3 \pm 0.9)$\% by comparing $\Bp\to \jpsi \Kp$ and $\Bm\to\jpsi \Km$ decays \cite{Aaij:2012jw}. A small subset of this background is from $\Lb^0$ decays, whose production asymmetry is not well known, $a_{\rm P}=(-1.0\pm4.0)$\%, but is consistent with zero \cite{Chatrchyan:2012xg}. The  $\Bz$ final states include  $\Dz$  and $\Dp$ hadrons, in proportions  determined according to the $\Dstarp/\Dp$ ratio in the measured exclusive final states.  In addition, we consider backgrounds coming from $\Bz,\Bp\to \Dsm K\mup$ decays, that provide a background asymmetry with opposite sign. 
We estimate this  background asymmetry to be (+0.01$\pm$0.04)\%.  The systematic uncertainty includes the limited knowledge of the inclusive branching fraction of the $b$-hadrons, uncertainties in the $b$-hadron production ratios, and in the charm semileptonic branching fractions, but is dominated by the uncertainty in the production asymmetry.
By combining these estimates,  we obtain $A_{\rm bkg}=(+0.05\pm 0.05)$\%.
%\clearpage
\section{Results}

We perform weighted averages of the corrected asymmetries  $A_\mu^c$ observed in each $p_{\rm T}\phi$ and $p_ x p_ y$ subsample, 
  using muon identification corrections both in the KS and MS sample (see Fig. 5).  In order to
cancel remaining detection asymmetry effects, the most appropriate way
to combine magnet up and magnet down data is with an arithmetic average \cite{Aaij:2012cy}. We then perform an arithmetic average of the four values of $A_\mu^c$ obtained with the two binning schemes chosen and with the two muon correction methods, assuming the results to be fully statistically correlated, and obtain $A_\mu^c=(+0.04\pm0.25)$\%.  The results are shown in 
  Table~\ref{tab:result summarynl0fix}. Finally, we correct for tracking efficiency asymmetries and background asymmetries,  and  obtain
$$\Ameas =(-0.03\pm 0.25 \pm 0.18)\%,$$
where the first uncertainty reflects statistical fluctuations in the signal yield and the second  reflects the systematic uncertainties. This gives
\begin{equation}
a^s_{\rm sl}=(-0.06\pm 0.50 \pm 0.36)\%.\nonumber
\end{equation}

%The asymmetry measured in the first momentum bin is slightly lower;  detailed studies of this momentum bin suggest this %to be consistent with a statistical fluctuation.

%%%%new summary figure Zhou
\begin{figure}[bt]
\begin{center}
\includegraphics[width=2.5 in]{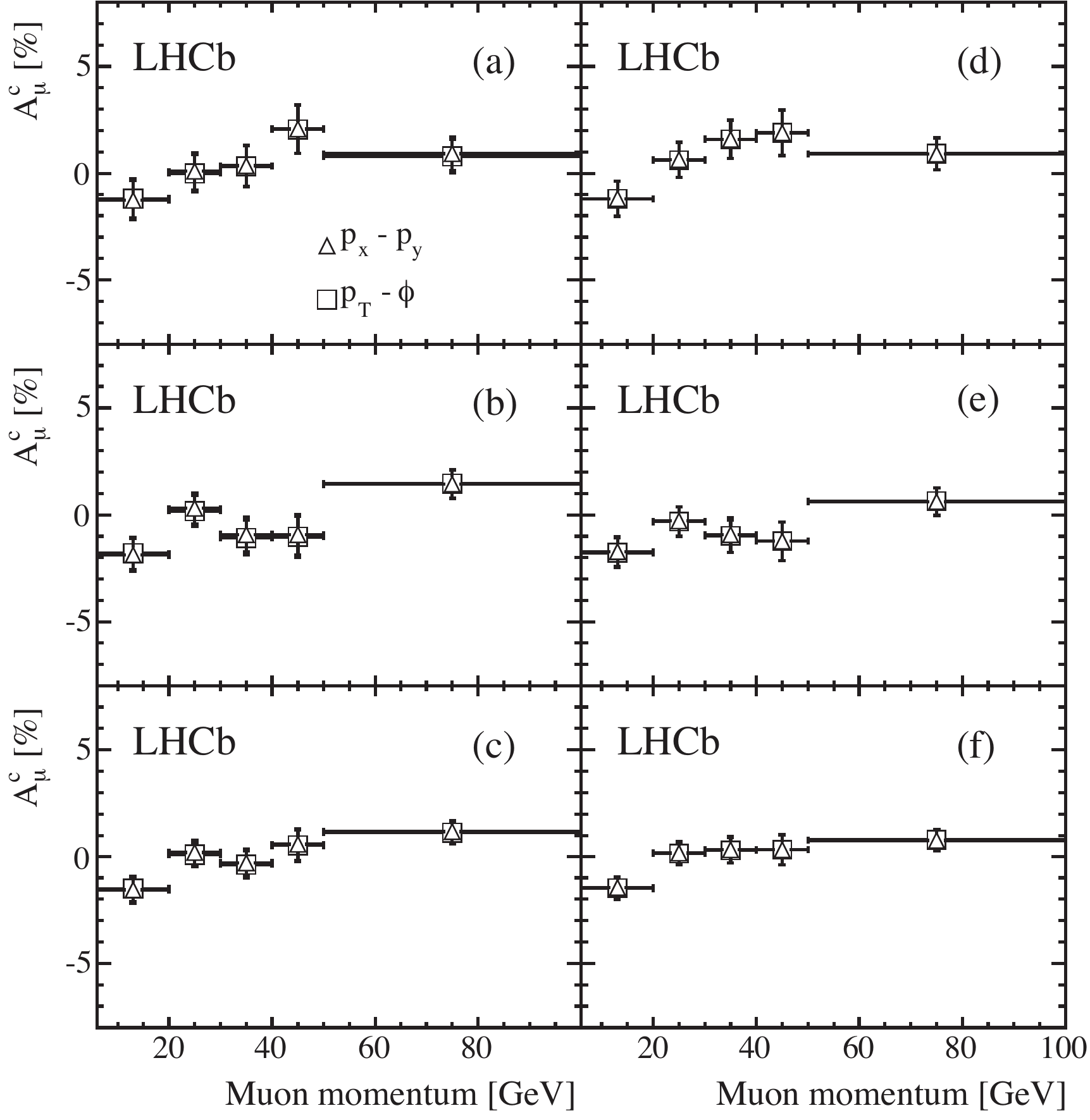}
\end{center}
\vspace{0pt}
\caption{\small Asymmetries corrected for relative muon efficiencies, $A_{\mu}^{\text{c}}$, examined in the five muon momentum intervals for  (a) magnet up data, (b) magnet down data and (c) average, using the KS muon calibration method. Then (d) magnet up data, (e) magnet down data and (f) average, using the MS muon calibration method in the two different binning scheme. }
\label{fig:fig21_A_corr_mu_phipi}
\end{figure}
%%%%%%%%%%

%%%%

\begin{table}[tb]
  \caption{\small Muon efficiency ratio corrected asymmetry $A_{\mu}^c$. The errors account for the statistical uncertainties in the  \Bs signal yields. }
  \begin{center}
  \resizebox{5.2in}{!}{
  \begin{tabular}{lccccc}

      \hline
      $A_{\mu}^c~~[\%]$        &  \multicolumn{2}{c}{KS muon correction} & \multicolumn{2}{c}{MS muon correction}   & Average\\ 
      \hline
      Magnet &\pxpy& \ptphi & \pxpy & \ptphi & ~~\\
      \hline
       Up            &  $+0.38 \pm 0.38$   & $+0.30\pm0.38$&  $+0.64\pm0.37$ & $+0.63\pm0.37$ & $+0.49\pm0.38$\\
       Down       &$-0.17\pm 0.32$  & $-0.25\pm0.32$&  $-0.60\pm0.32$  & $-0.62\pm0.32$ & $-0.41\pm0.32$ \\
      Avg.           &  $+0.11\pm0.25$      & $+0.02\pm0.25$&  $+0.02\pm0.24$  & $+0.01\pm0.24$ & $+0.04\pm0.25$\\\hline
     
  \end{tabular}}\end{center}
  \label{tab:result summarynl0fix}
\end{table}

We consider several sources of systematic uncertainties on $\Ameas$ that are summarized in Table~\ref{tab:systematic uncertainty}. 
 By examining the variations on the average $A_{\mu}^c$ obtained with different procedures, we assign a 0.07\% uncertainty, reflecting three almost equal components: the fitting procedure, the kinematic binning and a residual systematic uncertainty related to the muon efficiency ratio calculation. We study the effect of the fitting procedure by comparing results obtained with different models for signal and background shapes. In addition, we consider the effects of the statistical uncertainties of  the efficiency ratios, assigning 0.08\%, which is obtained by propagating the uncertainties in the average $A_\mu^{\rm c}$. The uncertainties affecting the background estimates are discussed in Sec.~5.
Possible changes in detector
acceptance during magnet up and magnet down data taking periods are estimated to
contribute 0.01\%. 
The software trigger systematic uncertainty is mainly due to the topological trigger algorithm and is estimated to be 0.05\%. These uncertainties are considered uncorrelated and added in quadrature to obtain the  total systematic uncertainty.

%summarized in Table~\ref{tab:systematic uncertainty}. The measured value of $\Ameas$ gives 

%%
%%
%%
%\clearpage
%\section{Systematic Uncertainties}
\begin{table}[b]
 \caption{\small Sources of systematic uncertainty on $\Ameas$.}
 \begin{center}\begin{tabular}{lc}\hline
     Source & $\sigma(A_{\text{meas}})$[\%]      \\\hline
%     \hline
     Signal modelling and muon correction &   0.07   \\
          Statistical uncertainty on the efficiency ratios & 0.08 \\
%     Background modeling in \Ds mass fit &       0.06              \\
       Background asymmetry & 0.05 \\
%     \hline
     Asymmetry in track reconstruction & 0.13 \\
     %     \hline
     Field-up and field-down  run conditions & 0.01 \\
%     \hline

     Software trigger bias  (topological trigger) & 0.05 \\
%     \hline
\hline
     Total  & 0.18 \\\hline
\end{tabular}\end{center}
 \label{tab:systematic uncertainty}
\end{table}

\clearpage

\section{Conclusions}
\begin{figure}[bt]
\begin{center}
\includegraphics[width=2.5 in]{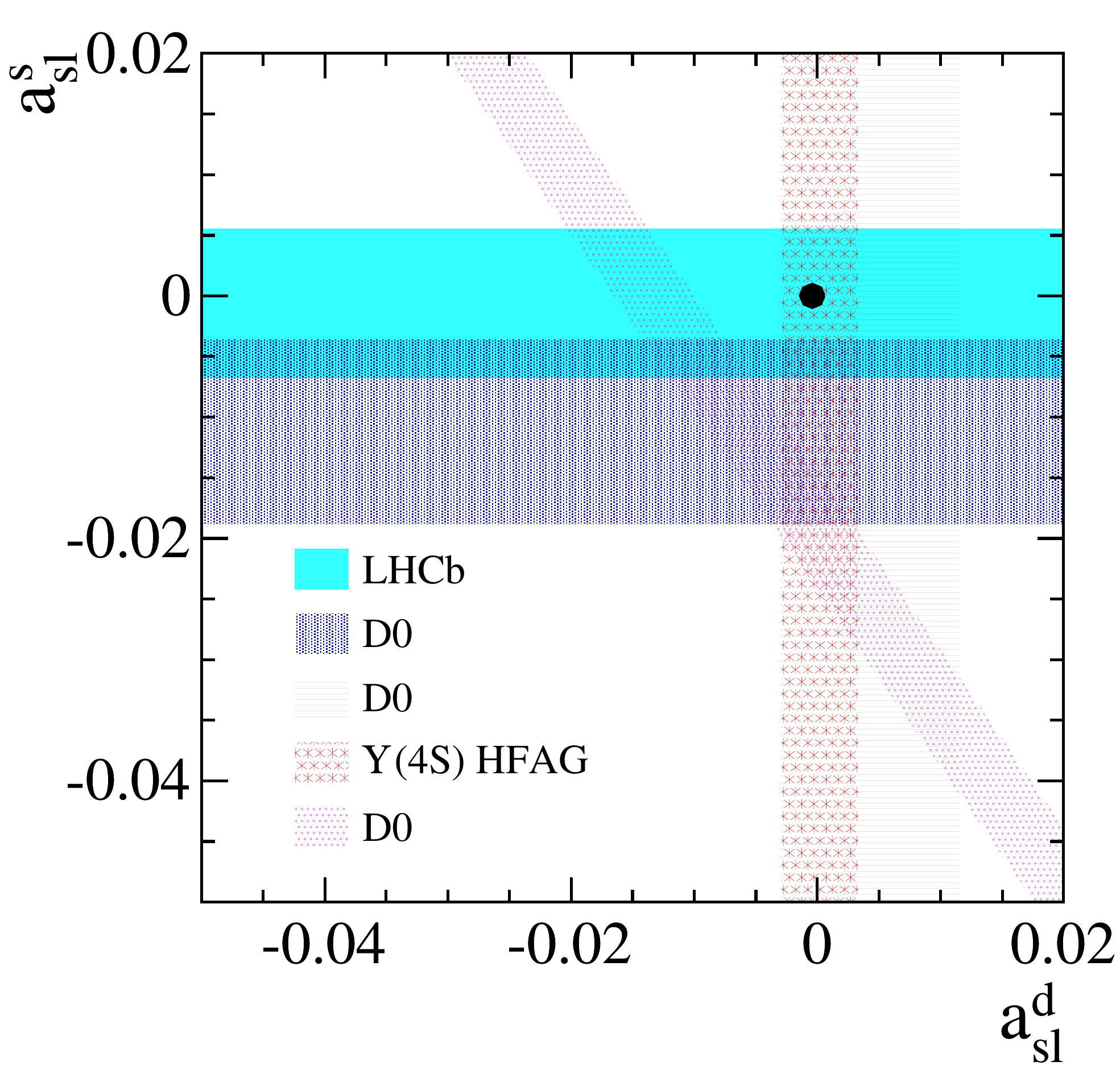}
\end{center}
\vspace{-3mm}
\caption{\small Measurements of semileptonic decay asymmetries. The bands correspond to the central values $\pm$1 standard deviation uncertainties, defined as the sum in quadrature of the statistical and systematic errors. The solid dot  indicates the SM prediction.} \label{D0-2-5}
\end{figure}

We measure the asymmetry $a^s_{\rm sl}$, which is twice the measured asymmetry between
$D_s^-\mu^+$ and $D_s^+\mu^-$ yields, to be
\begin{equation}
a^s_{\rm sl}=(-0.06\pm 0.50 \pm 0.36)\%. \nonumber
\end{equation}
%the recent D0 results on $a^s_{\rm sl}$ \cite{Abazov:2012zz} and $a^d_{\rm sl}$  \cite{Abazov:2012uia}, based on $D_{(s)}^{\pm}\mu^{\mp}$ events in a 10.4\,\invfb sample that give $a_{\rm sl}^s=(-1.12 \pm 0.74\pm 0.17)$\% and $a_{\rm sl}^d=(0.68 \pm 0.45\pm 0.14)$\%, and the average value of $a_{\rm sl}^d$ from $\Upsilon(4S)$ measurements  of $(0.02\pm0.31)$\% \cite{hfag}. In conclusion, our \asl ~result is the most precise determination to date, and is in agreement with the Standard Model prediction, thus we do not confirm the anomalous asymmetry reported by D0.
Figure~\ref{D0-2-5}  shows this measurement, the \dzero  measured asymmetries in dimuon decays in 1.96 TeV $p\overline{p}$ collisions
of  $A_{\rm sl}^b=(-0.787\pm 0.172\pm 0.093)$\% \cite{Abazov:2011yk},  $a^{d}_{\rm sl}= (0.68\pm0.45\pm0.14)\%$ \cite{Abazov:2012hha}, and $a^{s}_{\rm sl}=(-1.12\pm 0.74 \pm0.17)\%$ \cite{Abazov:2012zz}, and  the most recent average from $B$-factories \cite{Asner:2010qj},  namely $a_{\rm sl}^d=(0.02\pm0.31)$\%.
 Our result for $a^s_{\rm sl}$ is currently the most precise measurement made and is consistent with the SM.

\section*{Acknowledgements}

\noindent We express our gratitude to our colleagues in the CERN
accelerator departments for the excellent performance of the LHC. We
thank the technical and administrative staff at the LHCb
institutes. We acknowledge support from CERN and from the national
agencies: CAPES, CNPq, FAPERJ and FINEP (Brazil); NSFC (China);
CNRS/IN2P3 and Region Auvergne (France); BMBF, DFG, HGF and MPG
(Germany); SFI (Ireland); INFN (Italy); FOM and NWO (The Netherlands);
SCSR (Poland); MEN/IFA (Romania); MinES, Rosatom, RFBR and NRC
``Kurchatov Institute'' (Russia); MinECo, XuntaGal and GENCAT (Spain);
SNSF and SER (Switzerland); NAS Ukraine (Ukraine); STFC (United
Kingdom); NSF (USA). We also acknowledge the support received from the
ERC under FP7. The Tier1 computing centres are supported by IN2P3
(France), KIT and BMBF (Germany), INFN (Italy), NWO and SURF (The
Netherlands), PIC (Spain), GridPP (United Kingdom). We are thankful
for the computing resources put at our disposal by Yandex LLC
(Russia), as well as to the communities behind the multiple open
source software packages that we depend on.
  
\ifx\mcitethebibliography\mciteundefinedmacro
\PackageError{LHCb.bst}{mciteplus.sty has not been loaded}
{This bibstyle requires the use of the mciteplus package.}\fi
\providecommand{\href}[2]{#2}

\end{document}